\documentclass[preprint]{aastex}
\usepackage{spr-astr-addons}
\usepackage{url}
\usepackage{rotating}

\RequirePackage{color}

\begin{document}

\title{Minimum Spanning Tree cluster analysis of the LMC region above 10 GeV: detection of the SNRs N~49B and N~63A} 

%% Running heads
\shorttitle{MST cluster analysis of the LMC region above 10 GeV}
\shortauthors{R. Campana et al.}

%% Author and Affilations
\author{R.~Campana\altaffilmark{}}
\affil{INAF/OAS-Bologna, via Gobetti 93, I-40129, Bologna, Italy.} 
\and 
\author{E.~Massaro}
\affil{INAF/IAPS-Roma, via del Fosso del Cavaliere 100, I-00133, Roma, Italy}
\affil{In Unam Sapientiam, Roma, Italy}
\author{E.~Bernieri}
\affil{INFN/Sezione di Roma Tre, Roma, Italy}
\affil{Department of Mathematics and Physics, University of Roma Tre, Roma, Italy.}
\email{riccardo.campana@inaf.it} %% non-output

%% Alternate Affilations
%\altaffiltext{1}{Affilation}
%\altaffiltext{2}{}
%\altaffiltext{3}{}

\begin{abstract}
We present the results of a cluster search in the $\gamma$-ray sky images of 
the Large Magellanic Cloud region by means of the Minimum Spanning Tree 
algorithm at energies higher than 10 GeV, using 9 years of \emph{Fermi}-LAT data.
Several significant clusters were found, the majority of which associated with 
previously known $\gamma$-ray sources. 
New significant clusters associated with the supernova remnants N~49B and 
N~63A are also found, and confirmed with a Maximum Likelihood analysis of 
the \emph{Fermi}-LAT data.
\end{abstract}


\section{Introduction}\label{s:introduction}

The Large Magellanic Cloud (LMC) is a very interesting target for $\gamma$-ray astronomy
since the early suggestion by \cite{ginzburg72}, subsequently developed by \cite{ginzburg84}, 
who proposed to observe the Magellanic Clouds at high energies to test the hypothesis on the 
metagalactic origin of cosmic rays via the $\pi^0$ decay process.   
The $\gamma$-ray emission from LMC was discovered by EGRET-CGRO \citep{sreekumar92}.

The large amount of data collected by the \emph{Fermi}-Large Area Telescope (LAT) 
instrument \citep{ackermann12} in almost ten years of operation allows not only this type 
of studies but also the possibility to search for localized emission regions and 
point-like sources, that are particularly bright at energies in the GeV band and higher.
The  H.E.S.S. collaboration reported \citep{abramowski15}  observations of three 
sources up to about 10~TeV in the LMC, two of them associated with the Supernova
Remnants (SNR) N~157B \citep{abramowski12} and N~132D, the former containing 
the highest known spin-down luminosity pulsar PSR J0537$-$6910 
\citep{marshall98,cusumano98} with the fast period of 16 ms.

An analysis of the Fermi-LAT data of the first 6 years from the LMC was presented by
\cite{ackermann16} 
who detected four point sources (P1 to P4): one corresponding to the pulsar PSR J0540$-$6919
\citep{ackermann15a}, and two associated with the SNRs N~157B and N~132D,
while one source (P3) has been recently identified with the first extragalactic $\gamma$-ray binary 
within the SNR DEM L241 \citep{corbet16}.

In a series of papers \citep{bernieri13,paperI,paperII,paperIII,paperIV,campana17} we applied successfully 
the Minimum Spanning Tree (hereafter MST, \citealt{campana08,campana13}) source-detection 
method for searching new $\gamma$-ray sources closely associated with known BL Lac 
objects and blazar candidates, and illustrated how MST works in finding clusters having 
a small number of photons, but likely related to point-like sources.

In this paper we present the results of a MST analysis aimed at detecting photon clusters in 
the \emph{Fermi}-LAT data acquired in 9 years of observation above 10~GeV in a sky region 
containing the LMC, in order to identify candidate sources in the high energy $\gamma$-rays.
It should be emphasized that the dataset analyzed in this paper has a significant longer exposure than
the previous \emph{Fermi}-LAT studies \citep{ackermann16}, and employs the much improved
Pass~8 event processing \citep{atwood13}.

In particular, we found some evidence for possible extended structures and studied which 
cluster parameters can be used to discriminate them from patterns originated by pointlike 
sources.
Moreover, our results indicate the occurence of significant photon clustering close to 
the position of some bright SNRs not previously associated with $\gamma$-ray emitters. 

The outline of this paper is as follows. 
In Section~\ref{s:mst} the MST algorithm is briefly described, while the analysis 
of the LMC region is presented in Section~\ref{s:analysis}. 
In Section~\ref{s:snr} we deepen the analysis of some interesting SNR counterparts to 
MST clusters, while in Section~\ref{s:conclusions} the results are summarized.

\section{Photon cluster detection by means of the MST algorithm}\label{s:mst}

The MST is a topometric algorithm employed for searching clusters in a field of points. 
A brief description of the method was presented in \cite{paperI}, while a 
more detailed description with the statistical properties can be found in 
\cite{campana08,campana13}.
In this section the principal characteristics of the MST method are summarized for 
the sake of completeness. 

Considering a two-dimensional set of $N_n$ points, or \emph{nodes}, the set
$\{\lambda_i\}$ of weighted edges connecting them can be defined. 
The MST is the unique \emph{tree} (a graph without closed loops) connecting all the 
nodes with the minimum total weight, $\min [\Sigma_i \lambda_i]$. 
The edge weights, in the case of a region on the celestial sphere, are the angular distances 
between photon arrival directions. 

Once the MST is computed, a set of subtrees corresponding to clusters of
photons is extracted by means of the two following operations (\emph{primary} selection):
\emph{i) separation}: removing all the edges having a length 
$\lambda \geq \Lambda_\mathrm{cut}$, the separation value, that can be defined in units 
of the mean edge length $\Lambda_m = (\Sigma_i \lambda_i)/N_n$ in the MST, to obtain a 
set of disconnected sub-trees\footnote{Throughout this paper the simplyfing convention to write for instance 
$\Lambda_\mathrm{cut} = 0.7$ instead of $\Lambda_\mathrm{cut} = 0.7 \Lambda_m$ will be used.};
\emph{ii) elimination}: removing all the sub-trees having a number of nodes $N \leq N_\mathrm{cut}$, 
leaving only the clusters having a size over a properly fixed threshold. 
The remaining set of sub-trees provides a first list of candidate clusters. A \emph{secondary} 
selection must be applied for extracting the most robust candidates for $\gamma$-ray sources.
In \cite{campana13} a suitable parameter for this selection was introduced,  
 the so-called cluster \emph{magnitude}:
\begin{equation}
M_k = N_k g_k  
\end{equation}
where $N_k$ is the number of nodes in the cluster $k$ and the \emph{clustering parameter} $g_k$ 
is the ratio between $\Lambda_m$ and $\lambda_{m,k}$, the mean length over the $k$-th cluster edges. 
The probability to obtain a given magnitude value combines that of selecting a cluster with
$N_k$ nodes together with its ``clumpiness'' with respect to the mean separation in the field.  
On the basis of comparative tests performed in simulated and real \textit{Fermi}-LAT fields, 
\cite{campana13} found that $\sqrt{M}$ has a linear correlation with other statistical 
source significance parameters, e.g. those derived from a wavelet-based algorithm or Maximum Likelihood 
\citep{mattox96}
analysis, and therefore it can be a good estimator of statistical significance of MST clusters.
In particular, a lower threshold value of $M$ around 15--20 would reject the 
large majority of spurious low-significance clusters. 

Other characteristic parameters for the clusters can be then computed.
Examples are the centroid coordinates, obtained
by means of a weighted mean of the cluster photon arrival directions \citep[see][]{campana13} 
and the radius of the circle centred at the centroid and containing the 50\% of photons in the cluster, the 
\emph{median radius} $R_m$. The latter parameter, for a cluster that can be associated with a genuine pointlike 
$\gamma$-ray source, should be smaller than or comparable to the 68\% containment radius of 
instrumental Point Spread Function (PSF).
This radius varies from 0\fdg2 at 3 GeV to 0\fdg14 at 10 GeV in the case of a bright 
source for diffuse class front-observed events \citep{ackermann13a}.
Moreover, it can also be expected that the angular distance beetween the positions of the cluster centroid 
and the possible optical counterpart are lower than the latter value.
Another useful parameter for understanding the structure of a cluster is its \emph{maximum radius}
$R_\mathrm{max}$, defined as the distance between the centroid and the farthest photon in the cluster,
which is expected to be of some arcminutes for a point-like source and a few tens of arcminutes either for extended
structures or for unresolved close pairs of sources.

\begin{figure*}[ht]
\centering
\includegraphics[width=0.75\textwidth]{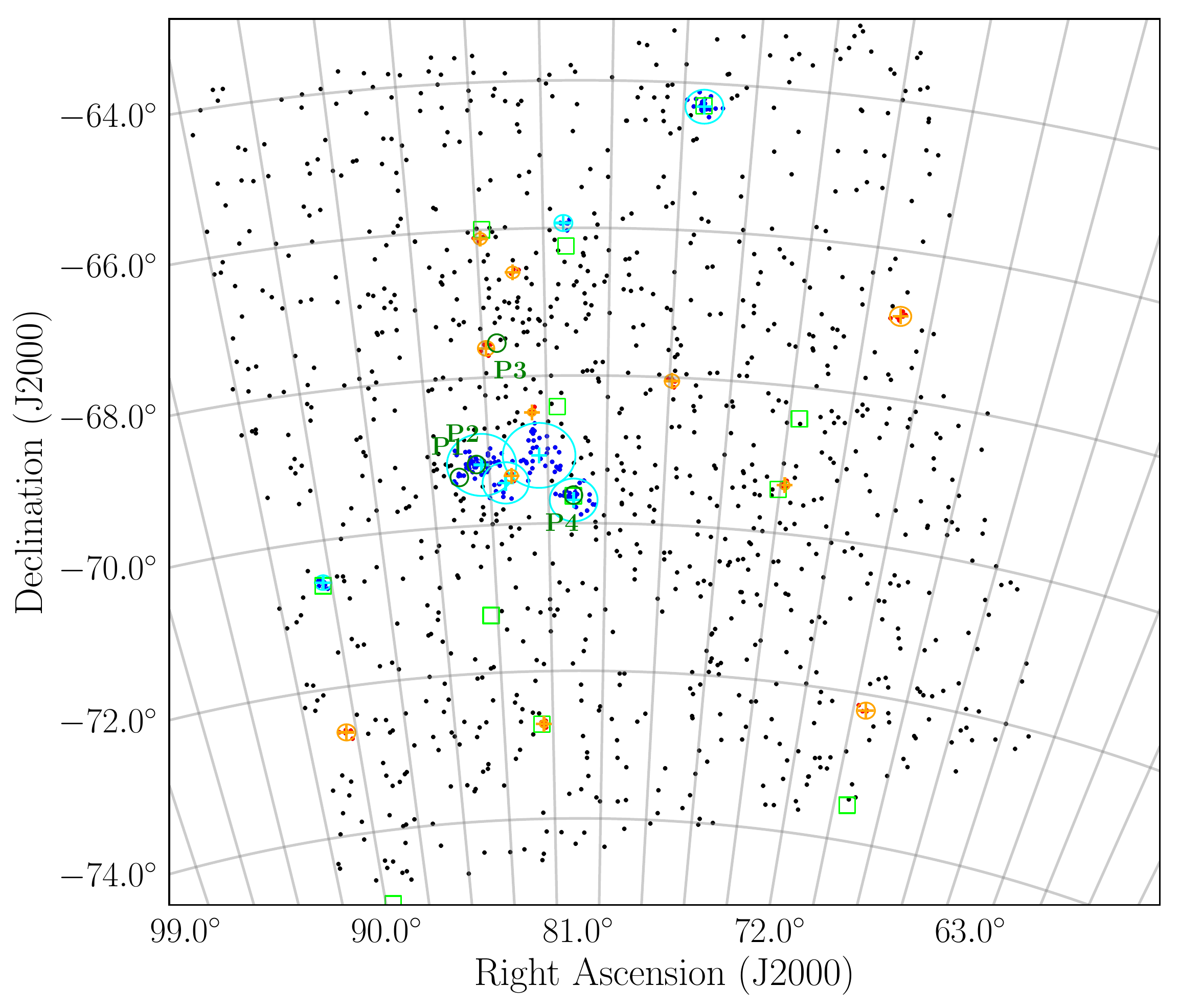}
\caption{Photon map in equatorial coordinates of the sky region (12\degr$\times$9\degr\ in a Galactic coordinate frame) centered at 
LMC at energies higher than 10~GeV. 
Blue and red points are the photons corresponding to the high and low significance clusters in Tables 1 and 2.
Symbol codes are: position of the MST high significance clusters above 10 GeV (cyan crosses), MST 
low significance clusters (orange crosses), 3FGL sources (light green open squares), 
$\gamma$-ray point sources P1--P4 found by \cite{ackermann16} (open green circles).
The cyan and orange circles are centered on the MST centroids and have a radius equal to $R_\mathrm{max}$.}
\label{skymap_f1}
\end{figure*}

\section{Cluster analysis of LMC high energy $\gamma$-ray data}\label{s:analysis}

LAT data (Pass 8R2) above 10~GeV, covering the whole 
sky in the 9.0 years time range from the start of mission (2008 August 04) up to 2017 
August 04, were downloaded from the FSSC 
archive\footnote{\url{http://fermi.gsfc.nasa.gov/ssc/data/access/}}.
Standard cuts on the zenith angle, data quality and good time intervals were applied.
Then, a region 12\degr$\times$9\degr\ approximately centered at the LMC was selected 
and the MST algorithm was applied. 
Figure~\ref{skymap_f1} shows the photon map in this region at energies higher than 10 GeV.

MST was first applied to the $>$10~GeV sky with $\Lambda_\mathrm{cut} = 0.7$, and $N_\mathrm{cut} = 3$, 
while the secondary selection was used to select significant concentrations only.
The choice of the optimal $\Lambda_\mathrm{cut}$ is not a simple task, because of the presence of
a diffuse and non-homogeneous emission in the LMC \citep{ackermann16} which can make difficult to resolve 
nearby clusters. 
We thus performed other analyses with shorter $\Lambda_\mathrm{cut}$ values to compare the results 
with those of the longer separation length.

The secondary selection was performed by adopting different thresholds for different cluster 
sizes (see Table~\ref{t:selection}  for a summary).
For clusters having a number of photons $N > 5$, a threshold of $M \ge 20$ was adopted,
while for clusters having $N = 4$ and $N = 5$ nodes, a threshold of $g \ge 3.5$ and 
$g \ge 3.0$ (corresponding to $M\ge14$ and $M\ge15$, respectively) was chosen in order 
to extract only structures with a photon density much higher than the surrounding and reduce 
in this way the probability for random clustering, as suggested by numerical simulations 
\citep{campana13}. 
We also extended our analysis to low significance clusters ($M < 20$), because of the 
possible associations either with extended features, or with faint sources, as verified 
by \emph{a posteriori} comparisons with literature data sets.
According to \citet{campana13} we expect that a percentage of around 50\% of these low
significance and ``poor'' clusters could be spurious and therefore those expected to be
associated with an interesting counterparts must be validated by further analysis.

\begin{table}[ht]
\caption{Criteria for the secondary selection}\label{t:selection}
\centering
\begin{tabular}{cc}
\hline
Cluster size & Selection criterium \\
\hline
$N>5$ & $M\ge20$\\
$N=5$ & $g\ge3.5$ \\
$N=4$ & $g\ge3.0$\\
\hline
\end{tabular}
\end{table}%

\begin{table*}
\caption{Coordinates and main properties of MST clusters with $M > 20$ detected in the LMC 
sky region at energies higher than 10 GeV. 
Celestial coordinates are J2000, angular distances $\Delta \theta$
are computed between the centroids of MST clusters and those of indicated counterparts.
The first section reports clusters detected using $\Lambda_\mathrm{cut} =  0.7$,
while the lower section those found with $\Lambda_\mathrm{cut} =  0.5$.
The letter ``e'' indicates likely extended structures, see main text for details.
}
\centering
{\small
\begin{tabular}{rrrrrccrrcl}
\hline
RA~~~ & Dec ~~~& $l$~~~ & $b$~~~ & $N$ & $g$~~~ & $M$~~~ & $R_m$ & $R_\mathrm{max}$ & $\Delta \theta$ &  Possible \\
\degr~~ & \degr~~  & \degr~~  & \degr~~  &     &        &        & \arcmin~   &  \arcmin~      &   \arcmin~   &  counterparts\\
\hline
\hline
~~~~$\Lambda_\mathrm{cut} =  0.7$  &    &         &         &    &        &         &       &     &                 &  \\
\hline
  77.427 & $-$64.315 & 274.307 & $-$35.210 & 20 &  3.129 &  62.576 &  4.4 & 13.6 & 2.4 &  FL8Y J0510.0$-$6417  \\ 
  81.124 & $-$69.686 & 280.365 & $-$32.823 & 26 &  3.273 &  85.099 &  8.7 & 17.4 & 4.1 &  P4 \\  
         &           &         &           &    &        &         &      &      & 1.7 &  N132D  \\
  81.395 & $-$65.934 & 275.927 & $-$33.308 &  6 &  3.516 &  21.094 &  2.8 & 6.4  & 3.4 &  N~49B    \\
         &         &           &           &    &        &         &      &      & 9.2 &  N49   \\
e 82.289 & $-$69.077 & 279.571 & $-$32.520 & 34 &  2.542 &  86.424 & 14.3 & 26.3 & 6.5 & FL8Y J0530.0-6900e  \\   
e 83.466 & $-$69.436 & 279.920 & $-$32.055 & 12 &  2.276 &  27.311 &  8.2 & 16.7 &   ---  & --- \\  
  84.258 & $-$69.178 & 279.575 & $-$31.810 & 57 &  4.200 & 239.427 &  9.0 & 25.1 & 4.0 & P2, N157B  \\ 
  90.292 & $-$70.566 & 280.993 & $-$29.633 & 11 &  5.793 &  63.719 &  2.2 & 6.0  & 2.6 & 5BZQJ0601-7036  \\    
\hline
\hline
~~~~$\Lambda_\mathrm{cut} =  0.5$  &    &         &         &    &        &         &       &     &                 &  \\
\hline
  77.435 & $-$64.318 & 274.310 & $-$35.209 & 16 &  3.999 &  63.988 &  3.2 & 10.0 & 2.3 &  FL8Y J0510.0$-$6417  \\ 
  81.231 & $-$69.618 & 280.279 & $-$32.798 & 17 &  4.008 &  68.133 &  4.8 & 15.2 & 2.1 &  P4 \\  
         &           &         &           &    &        &         &      &      & 1.7 &  N132D  \\
  81.399 & $-$65.933 & 275.923 & $-$33.308 &  5 &  4.585 &  22.925 &  2.2 & 5.4  & 3.3 &  N~49B    \\
         &           &         &           &    &        &         &      &      & 9.5 &  N49   \\
  81.663 & $-$69.251 & 279.816 & $-$32.711 &  6 &  3.523 &  21.137 &  2.3 & 6.4  & 6.1 &  SNR B0528-692  ? \\ 
  82.763 & $-$69.157 & 279.636 & $-$32.341 &  5 &  4.040 &  20.200 & 2.0  &  5.3 &  ---   & ---         \\ 
  84.228 & $-$69.174 & 279.573 & $-$31.822 & 46 &  4.913 & 225.979 & 7.3  & 16.0 & 5.1 & P2, N157B  \\ 
  85.003 & $-$69.303 & 279.686 & $-$31.534 &  4 &  5.404 &  21.614 & 1.6  &  2.5 & 1.9 & P1, PSR J0540$-$6919 \\  
  90.292 & $-$70.566 & 280.993 & $-$29.633 & 11 &  5.793 &  63.719 & 2.1  &  6.0 & 2.6 & 5BZQJ0601-7036  \\  
\hline
\end{tabular}
}
\label{t:rich}
\end{table*}

\begin{table*}
\caption{Coordinates and main properties of some low-significance MST clusters with $M < 20$ 
detected in the LMC sky region at energies higher than 10 GeV (except for the latest entry, 
found at energies higher than 7 GeV).
Celestial coordinates are J2000, angular distances $\Delta \theta$ are 
computed between the centroids of MST clusters and those of indicated 
counterparts.
See the main text for a discussion.}
\centering
{\small
\begin{tabular}{lrrrcrcrrrcl}
\hline
RA~~~ & Dec ~~~& $l$~~~ & $b$~~~ & $\Lambda_\mathrm{cut}$&$N$ & $g$~~~ & $M$~~~ & $R_m$ & $R_\mathrm{max}$ & $\Delta \theta$ &  Possible \\
\degr~~ & \degr~~  & \degr~~  & \degr~~  &  &    &        &        & \arcmin~   &  \arcmin~      &   \arcmin~   &  counterparts\\
\hline
\hline
  69.500 & $-$72.216 & 284.620 & $-$35.709 &  0.6  & 4 &  3.379 &  13.515 &  1.6 &  6.8 &   ---  & ---       \\   
  73.854 & $-$69.344 & 280.699 & $-$35.373 &  0.6  & 4 &  3.726 &  14.905 &  2.0 &  3.8 & 5.9 &   3FGL J0456.2$-$6924  \\ 
  82.373 & $-$72.715 & 283.801 & $-$31.866 &  0.7  & 4 &  3.539 &  14.156 &  2.5 &  3.2 & 2.3 &   FL8Y J0529.3$-$7245  \\  
  82.492 & $-$68.491 & 278.871 & $-$32.534 &  0.7  & 4 &  4.137 &  16.549 &  1.3 &  2.9 & 19. &   3FGL J0526.6$-$6825e \\ 
  82.952 & $-$66.590 & 276.607 & $-$32.605 &  0.5  & 4 &  4.077 &  16.307 &  1.9 &  4.7 & 3.6 &   FL8Y J0531.8$-$6639e \\  
  83.256 & $-$69.342 & 279.823 & $-$32.141 &  0.6  & 4 &  3.225 &  12.901 &  2.5 &  4.6 &  ---   &   possible satellite   \\
  83.889 & $-$66.119 & 276.006 & $-$32.281 &  0.6  & 4 &  3.482 &  13.926 &  2.3 &  4.7 & 5.0 &   N~63A                \\  
         &           &         &           &       &   &        &         &      &      & 7.5 &   3FGL J0535.3$-$6559  \\
\hline
\hline
  77.921 & $-$68.048 & 278.713 & $-$34.278 &  0.6  & 4 &  2.915 &  11.658 &  2.9 &  5.3 & 1.2 &   FL8Y J0511.5$-$6803  \\  
  83.888 & $-$67.605 & 277.753 & $-$32.129 &  0.7  & 4 &  1.849 &   7.397 &  4.5 &  6.2 & 2.8 &   P3, DEM L241  \\  
  90.406 & $-$72.613 & 283.343 & $-$29.505 &  0.7  & 5 &  2.378 &  11.889 &  3.5 &  6.7 & 2.0 &   FL8Y J0601.3$-$7238  \\ 
\hline
\hline
70.969  & $-$66.884 & 278.210 & $-$37.154 &  0.7  & 6 &  2.289 &  13.732 &  3.7 &  7.5 & 3.2 &   AllWISE CRATES \\ 
\hline
\end{tabular}
}
\label{t:poor}
\end{table*}

\subsection{High significance clusters}

The photon map in Figure~\ref{skymap_f1} shows a sky region containing the LMC where
the clusters selected by the MST analysis are reported together with other sources
from the literature.
It is apparent that the photon density is not homogeneous and that there
is a higher concentration approximately 
around $RA = 81\degr$, $Dec = -69\degr$,
which includes the well known 30 Doradus complex \citep{foreman15}.
In this region several contiguous clusters are found, marked by blue and red dots in the figure.

Centroid coordinates and other main parameters of the high significance 
($M \geq 20$) clusters selected with $\Lambda_\mathrm{cut} =$ 0.7 are reported in 
the first part of Table~\ref{t:rich}, while in the second part are those 
found with $\Lambda_\mathrm{cut} = 0.5$.
In the former part there are 7 clusters (see also Figure~\ref{skymap_f1}), five of 
which have a positional correspondence with sources in the 3FGL catalogue 
\citep{acero15} and with the sources found by \cite{ackermann16}, which are 
indicated by P2 and P4.                                        
In the following we describe some of their properties and possible associations.

\paragraph{MST(77.427, $-$64.315).}
This cluster is spatially close to the 3FGL J0509.7-6418, also reported as 1FHL J0509.9-6419, 
and a possible $\gamma$-ray source included in the preliminary 8-years source 
list\footnote{\url{https://fermi.gsfc.nasa.gov/ssc/data/access/lat/fl8y/}. 
Note that this catalog by no means official and will be soon superseded by the 
future 4FGL catalog.} 
provided by the \emph{Fermi}-LAT collaboration (also known as FL8Y), FL8Y J0510.0$-$6417, which does 
not have a well established counterpart: no known SNR is in close proximity, whereas it is 
associated with the bright X-ray source RBS 0625 \citep{schwope00}. 
This latter source has been classified as a quasar in the XMM-Newton Slew Survey \citep{warwick12}, 
and an interesting possibility is that it could be one of the background AGNs found in LMC.

\paragraph{MST(81.124, $-$69.686).}
This rich cluster is closely associated with the other SNR N132D (source P4 in 
\citealt{ackermann16}) that is at an angular distance from the centroid less than 2\arcmin.

\paragraph{MST(82.289, $-$69.077).}
This cluster has a low $g$ and the highest $R_\mathrm{max}$ and $R_m$ values; on the basis of these values it is likely to be an
extended structure (as indicated by the note ``e'' in Table~\ref{t:rich}), and appears directly associated with the source 
FL8Y J0530.0$-$6900e, also reported as extended in the preliminary \emph{Fermi}-LAT list.
Moreover, it is close to the other low $g$ cluster MST(83.466, $-$69.436) and the sum of their $R_\mathrm{max}$,
equal to 43\arcmin, is larger than the angular separation between their centroids (33\arcmin), indicating that they
can belong to a unique structure. 
Finally, note that the cluster centroid is moreover located between the two extended sources E1 and E3 reported by 
\cite{ackermann16}.

\paragraph{MST(84.258, $-$69.178).}
This is the richest cluster and is clearly associated with the SNR N157B 
(P2 source): the distance between the cluster centroid and the remnant is 4\arcmin, 
compatible with instrumental PSF at these energies.
However, the extension of this cluster is unusually large for a cluster having 57
photons, reaching a maximum radius $R_\mathrm{max}$ of 25\arcmin, much larger than the 
X-ray size 
of the PWN, smaller than 0\farcm5 \citep{wang01}.
The analysis performed applying $\Lambda_\mathrm{cut} = 0.5$ (Table~\ref{t:rich}, lower section) 
gives again a rich cluster with 46 photons but a higher clustering factor and a shorter
$R_\mathrm{max}=16\arcmin$, closer to the PSF size.
Considering that this cluster is embedded in the major photon concentration it is
likely that its extension can be originated by the aggregation of photons in the
field, thus a more realistic structure of this cluster is that of a point-like
structure embedded in an extended emission. 

\paragraph{MST(90.292, $-$70.566).}
This cluster can be the $\gamma$-ray counterpart of the blazar 5BZQ J0601$-$7036 
(using the naming of the 5th Edition of the Roma-BZCAT of blazars, \citealt{massaro14,massaro15}, 
also known as PKS~0601$-$70), corresponding to 3FGL J0601.2$-$7036. It is also reported in the FL8Y list. 

\paragraph{}
There are two high significance clusters not associated to previously reported 
$\gamma$-ray sources, which show interesting peculiarities.

\paragraph{MST(81.395, $-$65.934).}
This cluster has only 6 photons and a high $g$ indicating a compact structure with a maximum 
radius of about 6\arcmin\ as expected for a point-like source. 
It is at angular distance of 3\farcm4 from the SNR N~49B, while the nearby SNR N49 is at
9\farcm2, both undetected by \cite{ackermann16}.
Considering its $R_\mathrm{max}$ the latter association appears quite unlikely.
The analysis applying $\Lambda_\mathrm{cut} = 0.5$ (Table~\ref{t:rich}, lower section) 
substantially confirms this finding with a slightly reduced photon number but an 
increased $g$, thus giving the cluster an even higher magnitude value.
The properties of this cluster and his association with the SNR N~49B will be
discussed in detail in the following Section~\ref{s:snr}.

\paragraph{MST(83.466, $-$69.436).}
This is another structure with a low $g$ and a large $R_\mathrm{max}$, 
inside the region with the highest photon concentration in the field, suggesting that it might
be an extended feature, as previously observed for the cluster MST(82.289, $-$69.077).
To better understand its spatial structure, an analysis of the
region with shorter values of $\Lambda_\mathrm{cut}$ is useful.                                                    
With a separation length of 0.6 the cluster is divided in two not significant 
structures, one of which disappears when the separation length is reduced to 0.5.
Thus the extended structure appears confirmed.
There is no interesting counterpart, in particular no SNR in the 
\cite{maggi16} and \cite{bozzetto17} lists is close to this position.
An interesting possibility that should be further investigated is that the extended 
cluster is originated by some weak point-like source embedded in a diffuse emission.
For instance, the background quasar 
MQS J053242.46$-$692612.2
for which a spectroscopic redshift of 0.059 
is reported \citep{kozlowski13} and at about 6\arcmin\ from the centroid,
might be considered a possible candidate if it would exhibit a blazar nature.
In fact, its classification appear somewhat uncertain because it is undetected in the 
radio band, but there is a corresponding source in the 1RXH, WGACAT \citep{white00},
XMM and INTREFCAT \citep{ebisawa03} catalogues, although it was early classified as Low Mass X-Ray
Binary by \citet{liu01}.

\paragraph{}
The results of cluster finding with $\Lambda_\mathrm{cut} = 0.5$ 
(Table~\ref{t:rich}, lower section) generally confirm the existence of all compact 
structures found with the larger separation length, with the addition of some new 
interesting clusters having a small number of photons but a $g$ high enough
to give a $M$ value above the acceptance threshold.
Their properties are described in the following.

\paragraph{MST(81.663, $-$69.251).}
This cluster does not have any correspondence with any previously reported $\gamma$-ray sources.
It is located near to the SNR B0528-692, whose X-ray diameter is 3\farcm3 \citep{maggi16},
but the angular distance comparable to $R_\mathrm{max}$ weakens the association;
moreover, this SNR does not exhibit high brightness or other interesting properties,
suggesting it as a powerful $\gamma$-ray emitter.  

\paragraph{MST(85.003, $-$69.303).}
This cluster of only 4 photons has an unusual high $g$ and it is at a distance 
of only 1\farcm9 to the young pulsar PSR J0540$-$6919, already reported as source 
P1 by \cite{ackermann16}.
A cluster of 7 photons and $g = 4.33$ is found at 3\arcmin\ from the pulsar 
using an intermediate length $\Lambda_\mathrm{cut} = 0.6$, but no cluster is sorted 
out with longer separation values, because it is dispersed in the local background.
The proposed very likely association with PSR J0540$-$6919 has a rather soft
spectrum, and could explain this marginal detection.

\paragraph{MST(82.763, $-$69.157).}
This cluster corresponds to a cluster of 9 photons and a low $g = 2.562$ with 
$\Lambda_\mathrm{cut} = 0.6$, without any interesting counterpart. 
It can correspond to the extended sources E1 and E2 reported by \cite{ackermann16}.

\subsection{Low significance clusters}
Several other clusters were not considered in the analysis carried out up to now, because their magnitude $M$
was below the acceptance threshold and therefore the probability that they are random 
photon concentrations rather than features tracing genuine sources is generally higher than
for the previous clusters.
For a more complete description of the $\gamma$-ray emission structure of LMC, however,
it is interesting to extend our analysis to these clusters.
Moreover, we searched also if among other clusters sorted out by the MST there are some
associations to sources already reported in the literature and mainly 
detected at lower energies.
In first part of Table~\ref{t:poor} are reported 7 clusters having only 4 photons, $g \geq 3$ and,
consequently, $12 \leq M < 20$.
These clusters were found in the analyses with different $\Lambda_\mathrm{cut}$ and the 
reported parameters are those obtained with the highest $M$.
The clusters and their centroids are also plotted in Figure~\ref{skymap_f1}.
Note that four of these clusters are at a small angular distances from FL8Y (and also
3FGL) sources, indicating that these low $M$ structures, but with high $g$ values, must 
be taken into account for improving the completeness of search.

\paragraph{MST(82.492, $-$68.491).}
This cluster is in the neighborhood of the highest density photon 
region and could be associated with an extended 3FGL source.

\paragraph{MST(83.256, $-$69.342).}
It is located in the high density region and close to the boundary of the high significance 
cluster MST(82.289, $-$69.077) that can be associated with the source FL8Y J0530.0$-$6900e.   
A possibility to be taken into account is that it may be a \emph{satellite} of the larger 
feature that is not connected to it because the occurrence of an edge just longer the separation 
length.

\paragraph{MST(83.889, $-$66.119).}
It is the most interesting cluster in this list and a possible association with a near 3FGL source
cannot be excluded, albeit at a distance larger than $R_\mathrm{max}$;
its centroid is, however, closer to the powerful SNR N~63A, for which \cite{ackermann16} reported 
only an upper limit to its $\gamma$-ray flux.
A more complete analysis is given in the following Section~\ref{s:snr}.

\paragraph{}
For a more complete study on the possible sources in the LMC region we searched in our
cluster lists if there are some interesting structures close to sources already reported
in the literature and in other catalogues.
Despite their low significance above 10 GeV, we found three more clusters, listed
in the second part of the same Table~\ref{t:poor}, all close to FL8Y sources, two of which 
have $M$ just below the threshold, while the third one has quite low $g$ and $M$.
Their individual properties are given in the following.

\paragraph{MST(77.921, $-$68.048).}
This cluster is close to the source FL8Y J0511.5$-$6803 and its existence is confirmed by the analysis 
at energies higher than 5 GeV with $\Lambda_\mathrm{cut} = 0.7$ that gives a cluster of 
8 photons with $M = 18.262$ and an angular distance of only 1\farcm1 between their centroids.

\paragraph{MST(83.888, $-$67.605).}
This cluster was detected above 10 GeV despite the very low clustering degree and corresponds to the 
candidate high-mass X-ray binary (HMXB) CXOUJ053600.0$-$673507 \citep{corbet16},
located in the supernova remnant DEM L241 \citep{seward12}.
Moreover, it is close to the P3 source of \cite{ackermann16}.
The association of the $\gamma$-ray source with the X-ray binary is proved by the detection 
of a modulation of the $\gamma$-ray signal with the orbital period of 10.3 days, as measured 
from the radio, X-ray and radial velocity data \citep{corbet16}.
The MST analysis at energies higher than 3 GeV with $\Lambda_\mathrm{cut} = 0.7$ gives at
this location a cluster of 12 photons with $M = 26.171$, thus providing a satisfactory confirmation 
of our marginal high energy detection.

\paragraph{MST(90.406, $-$72.613).}
This cluster has 5 photons and is located at only 2\farcm0 from the FL8Y J0601.3$-$7238; it has a very 
likely counterpart in the blazar candidate WIBRaLS J060141.32$-$723833.2 \citep{dabrusco14} at 
an angular distance from the centroid of 1\farcm8.

\paragraph{}
Finally, no cluster was found above 10 GeV close to the flat spectrum radio source 
LMC B0443$-$6657, or CRATES J044318$-$665155 \citep{healey07}, recently detected in the LAT data 
by \cite{tang18}. 
It should be emphasized that the spectral analysis by \cite{tang18} gave above only upper limits 
above 10~GeV.
A MST cluster search at energies higher than 7 GeV gave a low significance cluster whose parameters 
are given in the last section of Table~\ref{t:poor}.
The CRATES source likely corresponds to an AGN \citep{secrest15} with a relatively bright mid-infrared 
counterpart (AllWISE J044318.24$-$665204.4) whose (undereddened) colours are well located in the 
FSRQ region of the \emph{WISE} Gamma-ray blazar strip \citep{dabrusco13,massaro16}.
On this basis, we conclude that the most likely counterpart of this source, if actually corresponding 
to that reported by \cite{tang18}, does not appear to be related to an LMC object but to a 
background FSRQ.

\section{The SNRs N~49B and N~63A: a more detailed analysis} \label{s:snr}

\begin{figure*}[ht]
\centering
\includegraphics[width=0.6\textwidth]{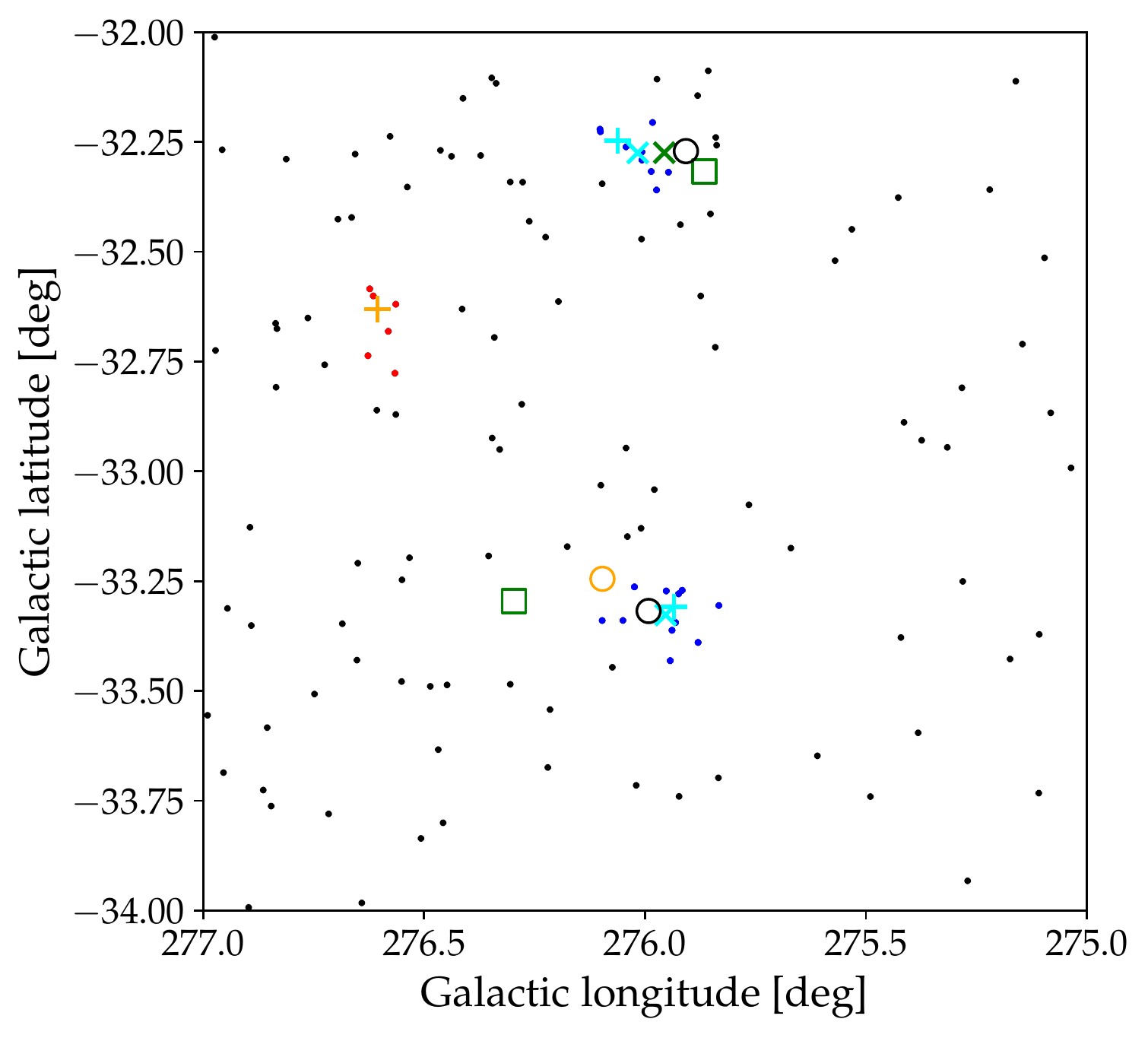}
\caption{Photon map in Galactic coordinates of the sky region ($2\degr\times2\degr$) around 
SNR N~49B and N~63A at energies higher than 6 GeV.
Photons are the black filled circles, while the blue ones correspond to 
the photons in the clusters found in the region with $\Lambda_\mathrm{cut} = 0.7$ and close to 
the SNRs (the upper one is around N~63A, while the lower one is around N~49B), 
and the red dots mark another low significance cluster. 
Cyan and orange pluses mark the weighted centroid positions, while cyan crosses are the unweighted centroid positions, 
and the green cross represents the centroid found for $E > 3$~GeV.
The positions of N~49B and N~63A are marked by black circles and N49 by an orange circle, 
respectively, while the green squares correspond to 3FGL sources.
}
\label{box_f2}
\end{figure*}

Up to now, the only two SNRs in LMC firmly detected in the $\gamma$-ray band are N~132D 
and N~157B \citep{ackermann16}, while the same authors reported upper 
limits for other interesting sources of the same class (cf. their Table 6).
Our MST analysis on the 9 year photon map above 10 GeV provided evidence for 
two clusters located very close to both the SNRs N~49B and N~63A.
These clusters, however, have a rather low number of photons and it is then 
necessary to extended the analysis to confirm these findings.

A new MST analysis was performed in a 6\degr$\times$6\degr\ subregion 
containing the two remnants, with different values of $\Lambda_\mathrm{cut}$ 
and extending the energy range down to 6~GeV.
These searches confirmed the existence of clusters close to both N~49B and N~63A.
Above 6~GeV a cluster at (81.3930, $-$65.9427), at the small angular separation of only 
2\farcm7 from the former SNR,  
is found with 11 photons and $M = 27.22$, thus well above the significance threshold,     
while another and richer cluster is found at (83.9622, $-$66.1698), at a distance
of 8\farcm0 from the latter SNR,   
with 9 photons and $M = 30.21$, $R_\mathrm{max}$ being $\sim$$8\farcm1$
(the angular distance reduces to 5\farcm6 using the \emph{unweighted} centroid).
Using a lower energy threshold of 3~GeV an even richer cluster of 19 photons at
(83.8927, $-$66.0977) with $M = 51.58$ is found and its separation from the SNR
decreases to 4\farcm6.
The existence of reliable high energy $\gamma$-ray source candidates is then 
fully confirmed and the association with the two SNRs appears more robust.
The photon map for $E > 6$~GeV of a $2\degr \times 2\degr$ region where
the two SNRs are located is shown in Figure~\ref{box_f2}. 
Extracting from the sample of \cite{maggi16} 
the SNRs with
a 0.3--8 keV X-ray luminosity brighter than $7 \cdot 10^{33}$~erg/s, 
only 3 are within this field, the two under consideration and N~49. 
In the plot, the cluster photons are in blue (high significance clusters) and red (low 
significance clusters), while the crosses mark their centroid coordinates: 
it is clear that N~49 (orange circle) is outside the photon concentration while
the association with N~49B is much more robust.

In this map it can be seen that both clusters have a pair of photons much 
closer between them than the other ones.
This particular occurrence for cluster with a low photon number can sligthly 
affect the estimates of the centroid coordinates, since they are weighted 
with the inverse of the squared edge length \citep[see][for a discussion]{campana13}.
Therefore, it is useful to consider the \emph{unweighted} centroids, also shown in
Figure~\ref{box_f2}, which result closer to the location of the SNRs.

\begin{figure*}[htb]
\centering
\includegraphics[width=0.6\textwidth]{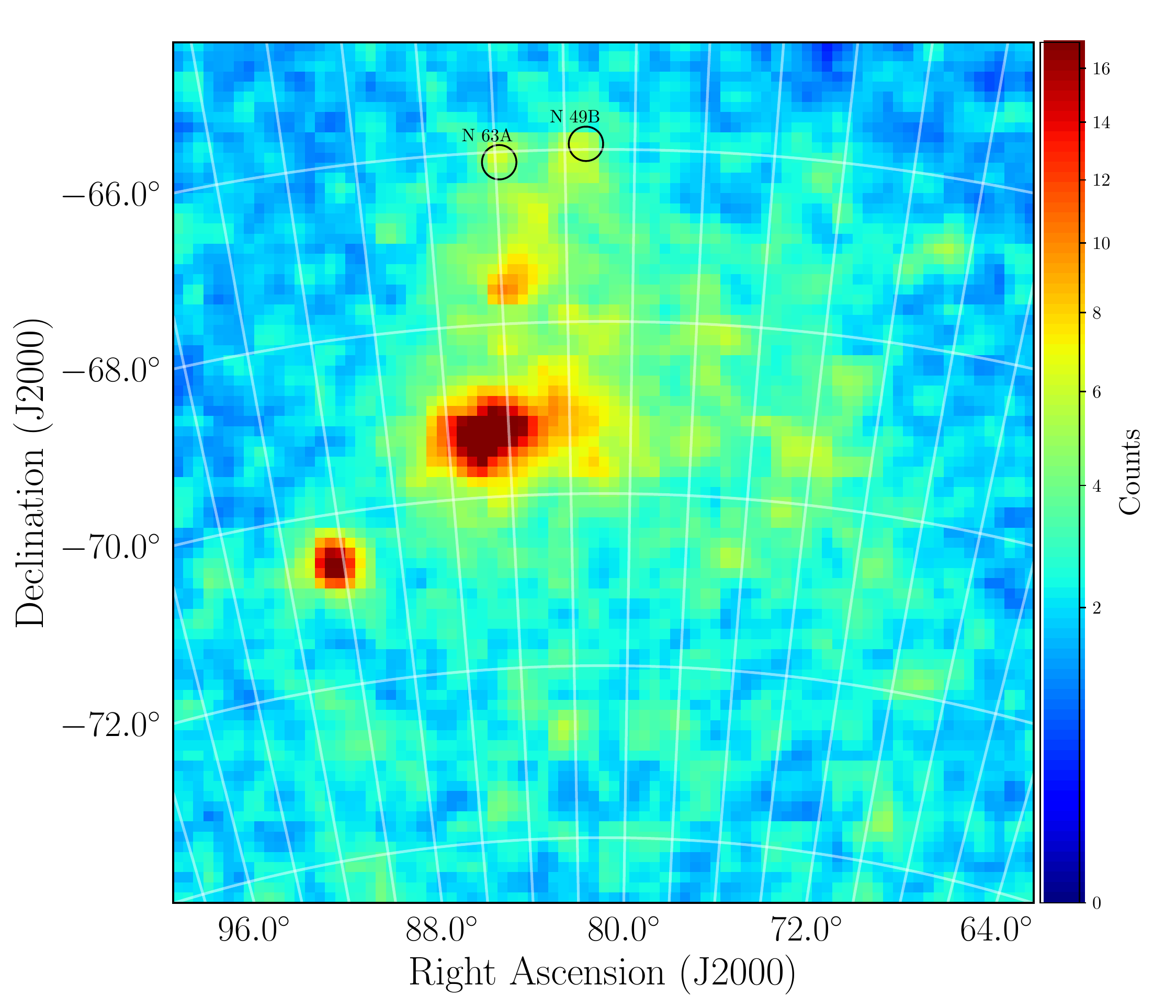}
\caption{Count map of the wide-scale LMC $\gamma$-ray emission at energies higher than 1~GeV
in equatorial coordinates, showing the locations of the two SNRs detected
by MST (black circles): N~49B and N~63A are located just above and below 
the declination circle $-66\degr$, respectively. Map with 0\fdg1 wide pixels,
smoothed with a Gaussian kernel one pixel wide based on 9-years \emph{Fermi}-LAT data, 
and with a square-root color scale and Aitoff projection.
}
\label{map_f3}
\end{figure*}

\subsection{Maximum Likelihood analysis}

To validate the existence of a source and to estimate its photon flux,
the standard binned Maximum Likelihood (ML) analysis was also performed at energies above 1~GeV in
a sky region of interest (RoI) with an extension of $10\degr \times 10\degr$ centered at
the equatorial coordinates $\alpha = 81\fdg4$, $\delta = -65\fdg9$.

In the case of LMC a major issue is the model for the local emission.
Therefore, three possible models were considered beside the isotropic and 
Galactic diffuse backgrounds:
$A$) 3FGL diffuse model for LMC (two 2D Gaussian spatial distributions for the full disk and 30
Dor emission\footnote{Template available at \url{https://fermi.gsfc.nasa.gov/ssc/data/access/lat/4yr_catalog/LAT_extended_sources_v15.tgz}.}) 
plus all the 3FGL point-like sources; $B$) the same model as $A$, but with the 3FGL
sources inside the RoI replaced by the four sources reported by \cite{ackermann16} (P1--P4);
$C$) like model $B$, but with the structured 4-component model derived by \cite{ackermann16} for the LMC diffuse
emission\footnote{Templates available at \url{https://fermi.gsfc.nasa.gov/ssc/data/access/lat/3FHL/LAT_extended_sources_v18.tgz}.}.

Spectral parameters for the background sources inside the RoI were fixed at their catalog values, with the exception of the flux normalisation.
Moreover, two point-like sources with a power law spectrum were added to the model, with a location given by the MST centroid coordinates, leaving both the
spectral index and the normalization as free parameters, in order to include the new MST-detected SNRs. 
Fit results are shown in Table~\ref{t:mlres}.

Model $A$ provided a significant detection for N~49B, 
with a test-statistics value of $\sqrt{TS} =  6.6$ and a 1--300~GeV flux of $(2.7\pm0.7)\cdot10^{-10}$ ph~cm$^{-2}$~s$^{-1}$.
A source at the MST coordinates for N~63A is not detected, although the very close 3FGL~J0535.3$-$6559 is detected with $\sqrt{TS} = 8.0$.
We can conclude that the latter 3FGL source could be the counterpart of the MST cluster.

With Model $B$ both sources are significantly detected, with $\sqrt{TS}$ of 7.9 and 9.0 for N~49B
and N~63A, respectively, with fluxes of $(3.5\pm0.7)\cdot10^{-10}$ and $(3.4\pm0.6)\cdot10^{-10}$ ph~cm$^{-2}$~s$^{-1}$.

With Model $C$ however the detections are less significant, with $\sqrt{TS}$ equal to 3.9 and 4.7,
because the local excesses in correspondence to the two remnants can be partially blended in 
the highly structured LMC diffuse emission.
Considering that the two SNRs are in a region of rather low diffuse local flux
(see the count map in Figure~\ref{map_f3}) and quite distant from the prominent region of 30 Doradus,
in particular at energies in the GeV band, the first two models are reasonably acceptable, 
and therefore ML results are not in conflict with our findings.

It is interesting to compare the Model $C$ results to the 95\% c.l. upper limits given by \cite{ackermann16}, Table 6, for both SNRs. Assumung a power law index of 2, their upper limits are $3.1\cdot10^{-7}$ and $4.8\cdot10^{-7}$ MeV cm$^{-2}$ s$^{-1}$ for N~49B and N~63A in 1--10 GeV, respectively.
Using the results in Table~\ref{t:mlres}, we obtain fluxes in the same band of $(7\pm3)\cdot10^{-7}$ and $(9\pm3)\cdot10^{-7}$  MeV cm$^{-2}$ s$^{-1}$, respectively, i.e. about a factor of two higher than the upper limits given in \cite{ackermann16}.

\begin{table*}[htb]
\caption{Maximum likelihood results. $\Gamma$ is the power law spectral index, and the flux is in the 1--300~GeV band. See main text for details.}\label{t:mlres}
\begin{center}
\begin{tabular}{ccccccc}
\hline
		& \multicolumn{3}{c}{N~49B} & \multicolumn{3}{c}{N~63A} \\
Model   & $\sqrt{TS}$ & $\Gamma$       & Flux   [ph cm$^{-2}$ s$^{-1}$]                   & $\sqrt{TS}$ & $\Gamma$       & Flux [ph cm$^{-2}$ s$^{-1}$] \\
\hline
A       & 6.6	     & $2.2 \pm 0.2$ & $(2.7\pm0.7)\cdot10^{-10}$ & ---        & ---            & --- \\
B       & 7.9	     & $2.3 \pm 0.2$ & $(3.5\pm0.7)\cdot10^{-10}$ & 9.0        & $2.1 \pm 0.2$ & $(3.4\pm0.6)\cdot10^{-10}$ \\
C       & 3.9	     & $2.0 \pm 0.3$ & $(1.3\pm0.6)\cdot10^{-10}$ & 4.7        & $1.9 \pm 0.2$ & $(1.5\pm0.5)\cdot10^{-10}$ \\
\hline
\end{tabular}
\end{center}
\end{table*}

\section{Summary and discussion}\label{s:conclusions}
The Large Magellanic Cloud contains a large population of young SNRs, including Pulsar 
Wind Nebulae and shell-like structures (superbubbles) produced by blast waves sweeping out 
the ambient interstellar medium \citep{maclow88}, well observed in 
the radio band and X-ray bands (see, for instance, the recent papers by 
\citealt{maggi16} and \citealt{bozzetto17}).
H.E.S.S. observations in the TeV band \citep{abramowski15} found the two 
powerful SNRs N~157B and N~132D, also detected by \cite{ackermann16} in the 
6 year \emph{Fermi}-LAT data.
The latter authors were also able to detect two more point-like sources (P1 and P3): one 
identified as PSR~J0540$-$6919, while the other one remained unassociated, and
reported only upper limits for other powerful SNRs expected to be possible high
energy emitters.

We analysed \emph{Fermi}-LAT sky images, obtained in 9 years of observations at energies 
higher than 10 GeV, where the photon density is rather low, allowing a robust detection 
of photon clusters with typical sizes comparable or larger than the instrumental PSF.
In the selection procedure  rather severe threshold values were adopted to reduce 
the possibility of spurious detections originated by random fluctuations of
background events.
The application of MST provided 7 high significance clusters (Table~\ref{t:rich}), five of them clearly 
associated with known $\gamma$-ray sources, while the remaning two are without a
correspondence in previous searches.
One cluster has a quite large angular radius, indicative of an extended emission, while
the other cluster, with only 6 photons but a rather high $g$, is found to have a
positional correspondence with the SNR N~49B, up to now undetected in the $\gamma$-ray 
band.
Moreover, a search performed using use of a shorter separation length $\Lambda_\mathrm{cut}$ 
provided three more clusters, one of which 
corresponding to the pulsar PSR J0540$-$6919 already reported by \cite{ackermann16},
and another one in close proximity of the SNR B0528$-$692.

For the sake of completeness, our investigation was also extended to other structures 
classified as low significance clusters, containing only 4 or 5 photons and having magnitude 
values $M < 20$ (Table~\ref{t:poor}).
Seven other clusters were selected, four of them associated to prevously known $\gamma$-ray 
sources. 
One of the three new clusters is located close to the SNR N~63A, and this association
is strengthened by the finding of high significance clusters in analyses at lower 
energies. 

The significance of the $\gamma$-ray emission from N~49B and N~63A was validated 
also by the ML analysis, with $\sqrt{TS}$ values higher than the usual detection 
threshold of 5, although this result is dependent on the precise model assumed for the background sources in the region of interest.

Finally, we searched for clusters likely associated with other literature sources and
found some structures with $M$ values too low to be accepted as confirmed detections.
A possible interpretation of these clusters is that they could be produced by very 
faint high energy sources, which above a few GeV would give only 2 or 3 photons in the 
considered timeframe, that grouped together with some background events are able to 
produce structures above the primary selection criteria.
It is interesting that clusters related to the source P3 by \cite{ackermann16} and
to two FL8Y sources are found, besides the \cite{tang18} source, whose likely counterpart appears
to be a background blazar.

Considering the above results, the number of SNRs and PWNe in the LMC detected in the $\gamma$-ray
band increases to four.
As a final remark, no cluster was found corresponding to the position of SN~1987A confirming 
that the high energy emission in the GeV band from this object, if present, is much 
fainter than from older remnants.

%______________________________________________________________
\begin{acknowledgements}
We are grateful to the anonymous referee, whose careful reading of the manuscript and his/her
useful comments greatly improved this paper.
We acknowledge the use of archival \emph{Fermi}-LAT data.
\end{acknowledgements}
%______________________________________________________________

\bibliographystyle{spr-mp-nameyear-cnd}
\bibliography{bibliography} % your references Yourfile.bib 

\end{document}